\documentclass[pra,showpacs,twocolumn,showkeys]{revtex4-1}
\usepackage{multirow}
\usepackage{amsmath}
\usepackage{amssymb}
\usepackage{graphics}
\usepackage{graphicx}
\usepackage{array}
\usepackage{color}
\usepackage[hypertex, colorlinks, colorlinks=false, linkcolor=blue, citecolor=green, filecolor=black, urlcolor=cyan]{hyperref}
\usepackage{ulem}

\begin{document}

\newcommand{\sinc}{{\rm sinc}}

\draft
\title{The influence of geometry and topology of quantum graphs on their nonlinear-optical properties}
\author{Rick Lytel, Shoresh Shafei, Julian H. Smith and Mark G. Kuzyk}
\affiliation{Department of Physics and Astronomy, Washington State University, Pullman,
Washington  99164-2814}

\date{\today}
\begin{abstract}
{\noindent We analyze the nonlinear optics of quasi one-dimensional quantum graphs and manipulate their topology and geometry to generate for the first time nonlinearities in a simple system approaching the fundamental limits of the first and second hyperpolarizabilities.  We explore a huge configuration space in order to determine whether the fundamental limits may be approached for specific topologies, independent of molecular details, when geometry is manipulated to maximize the intrinsic response. Changes in geometry result in smooth variations of the nonlinearities.  Topological changes between geometrically-similar systems cause profound changes in the nonlinear susceptibilities that include a discontinuity due to abrupt changes in the boundary conditions. We demonstrate the same universal scaling behavior for quantum graphs that is predicted for general quantum nonlinear optical systems near their fundamental limits, indicating that our results for quantum graphs may reflect general structure-property relationships for globally-optimized nonlinear optical systems. 
}
\end{abstract}
\pacs{42.65.An, 78.67.Lt}

\maketitle

\section{Introduction}

The search for materials with large electronic nonlinear optical susceptibilities has been underway for several decades due to their potential use in very high speed all-optical switching and modulation devices \cite{heebn99.01}, as well as for measurement of ultrafast phenomena in nonlinear optics \cite{brabe00.01}. At the turn of the century, the new theory of fundamental limits of the first and second hyperpolarizabilities in the off-resonance regime \cite{kuzyk00.01} showed that the best measured hyperpolarizabilities of molecules fell short of the limits by a factor of 30 \cite{Tripa04.01,tripa07.01}.  This surprising observation led to a search for quantum systems with potential energy functions that produce an optimized nonlinear response \cite{kuzyk06.02} and resulted in new concepts and realizations for structures with record hyperpolarizabilities \cite{zhou06.01,zhou07.02,perez07.01,perez09.01}, still far short of fundamental limits however. In fact, most reports of better molecules are simply seeing the effects of simple size scaling, not new physics.

While the hyperpolarizabilities of these numerically-optimized systems do not reach the limit, they share certain universal properties at their global maxima \cite{watkins09.01}.  A sampling of the entire solution space with Monte Carlo simulations that span all possible transition moments and energy spectra obeying nonrelativistic quantum mechanical sum rules revealed the existence of states and spectra that approach arbitrarily close to the fundamental limit \cite{kuzyk08.01,shafe10.01}.  An important fundamental question is whether real physical systems exist that approach the fundamental limits, and if so, what are the properties of their energy spectra that produce the larger responses \cite{shafe11.01}.   In hydrogen-like atoms, where the density of states increases with energy, all states with similar energies contribute equally to the nonlinearities, resulting in a diluted contribution to the nonlinear response.  A quantitative rule of thumb, the so-called Three-Level Ansatz, emerged from the theory of fundamental limits and states that only three states contribute for a system with a nonlinearity close to the limit, consistent with all observations and analysis to date. This strongly suggests that systems in which many states contribute yield low hyperpolarizabilities \cite{kuzyk00.01}, also consistent with all known data.  Thus, a promising path toward delineating structures with nonlinearities approaching their fundamental limits is to explore models with states and spectra mimicking three-level systems with large energy gaps and small transition moments of the higher-lying excited states \cite{shafe10.01} It is also notable that quantum wires have shown promise for attaining larger nonlinearities \cite{sande92.01,chen93.01,xia97.01, balle99.01} due to their confinement properties.  Intuitively, model systems exhibiting confinement, but with states and spectra having large energy gaps and small transition moments for higher-lying excited states, should be excellent vehicles for exploring the hyperpolarizabilities near the fundamental limits as the geometrical and topological features of the underlying structures are varied.

Quantum graphs are ideal model systems for such studies. A quantum graph is a network of metric edges and vertices supporting particle dynamics governed by a self-adjoint Hamiltonian that operates on the edges.  Boundary conditions are imposed at the vertices, including terminal points.  Quantum graphs have a complete set of eigenstates and an energy spectrum \cite{kotto97.01,berko06.01,kuchm04.01,kuchm08.01} describing dynamics on the graph.  An electron on the graph is assumed to be tightly bound in the transverse direction, yielding a quasi one-dimensional dynamical system. Quantum graphs thus possess appropriate spectra and states with which to explore structure-property relations of quantum-confined systems whose geometry and topology may be selected to generate responses that approach the fundamental limits.  Such models were first employed to explore properties of simple molecular structures \cite{pauli36.01}; but, we will invoke them in a detail-independent way to investigate general relationships among geometrical and topological properties of nonlinear optical systems as their response approaches the fundamental limits.

Quantum graphs with zero potential energy (bare edges) and nonzero potentials (dressed edges) have been solved using periodic orbit theory and extensively studied for their statistical properties and energy spectra \cite{kotto97.01,kotto99.02,blume02.01,blume02.02,dabag04.01,dabag02.01,dabag03.01}.  Interest in quantum graphs exploded in the mid-late 1990s when exact solutions for chaotic quantum graphs were discovered.  Quantum graph theory research continues to blossom as a field of pure and applied mathematics, with applications to chaos, fractals, and spectral theory.  Can quantum graphs with suitable geometries and specific topological features produce optical nonlinearities approaching the fundamental limit?  To answer this, aggregates of quantum wires into simple structures were recently studied for the first time for their nonlinear optical properties \cite{shafe12.01,lytel12.01}.  The results showed that simple quantum loop graphs could achieve modest responses and offered a number of insights into the quantum graph model and its universal properties.

It is the aim of this work to expand the investigation to complex graphs and to delineate the exact dependence of the first and second hyperpolarizabilities on the topological features and geometries of the graphs in order to identify a quantitative path forward for the synthesis of nanowires and molecular systems with nonlinearities approaching the fundamental limit \cite{zhou07.02}.  As noted above, nanostructures exhibit the desirable scaling with state number to achieve optimum response, and our results are presented in a scale-invariant way, ensuring that increases in nonlinearity are due to topological properties and not simply due to size scaling. The present work employs the simple model of electron dynamics on a quantum edge and shows that this model exhibits the universal properties expected from a quantum mechanical system satisfying the full sum rules, suggesting that it captures the essential physics required to describe the nonlinearities of quantum graphs.

To understand the origin of the nonlinear optical response in quantum graphs, the present investigation systematically investigates the role of geometry for graphs with the same topology, as well as the impact of changing topology with fixed geometry. We discover that geometric effects produce modest variations in the maximum response for a given topology, but that topological shifts actually control the dominant behavior of a class of geometrically-similar graphs.  This result implies that designers of molecular and nano-scale nonlinear optical structures should focus first on achieving the topological factors that produce the greatest response before optimizing their geometries (excluding those cases where geometric constraints might produce undesirable symmetries that suppress or eliminate nonlinear optical responses).

The optimum topologies are based upon the three-prong star graph, a key component in the assembly of larger graphical structures and of physical interest due to its exact spectral solution by a periodic orbit expansion \cite{pasto09.01}, its close connection to Seba billiards and Zeta functions \cite{harri11.01}, and many other problems \cite{brown09.01,berko99.01,ohya12.01,winn06.01}.  Our computation of the hyperpolarizabilities for a class of star graphs reveals the largest intrinsic response computed to date in any system, let alone an artificial material.

Table \ref{tab:results} displays the four sets of basic graphs studied in this paper.  The first set are bent wires with different geometries but linear topology.  The second set are the topologically-equivalent closed-loop graphs -- again with different geometries.  Below the double line, the next set are graphs with the same triangle geometry but which differ in topology.  The final set are 3-prong irregular (i.e.~irrational length ratios) star graphs, which are geometrically equivalent, but topologically distinct. In the Table, $\beta_{xxx}$ is the largest diagonal component of the first hyperpolarizability tensor, normalized to its fundamental limit, $|\beta|$ is its tensor norm, and $\gamma_{xxxx}$ is the largest diagonal component of the second hyperpolarizability, normalized to its maximum value.  The results displayed in the Table will be discussed later once the basic problem of solving a quantum graph for its nonlinear optical response is defined and solved.

The paper is organized as follows.  Section \ref{sec:NLOgraphs} discusses the solution of the Schr\"{o}dinger Equation for an electron inside a quantum graph.  Then, the basic machinery is developed for determining the tensors for the first hyperpolarizability, $\beta_{ijk}$, and the second hyperpolarizability, $\gamma_{ijkl}$ of a graph with specific geomety/topology.  The solution to a general graph requires computation of transition moments (position matrix elements) along sequential edges of a graph, in loops, and at intersections or stars, where several edges meet at a single vertex.  Section \ref{sec:NLOgraphs} shows how to calculate these for any graph.  We use the irreducible spherical tensor representation in Section \ref{sec:tensors} to extract all of the physical quantities from the transition moment calculations for each graph.  We then analyze the origins of the geometrical and topological effects. Section \ref{sec:results} presents our results for the hyperpolarizability tensors and reviews the details of the results displayed in Table \ref{tab:results}. This section also examines the scaling properties of the nonlinearities of the graphs as they approach their best values, revealing the universal behavior expected from a full quantum mechanical model of a nonlinear optical system.  The results are shown to support the thesis that the simple one-electron model meets all of the requirements to be classified as a realistic model of a physical quantum-confined system.  Finally, Section \ref{sec:Conclusion} concludes with a discussion about the future directions for expanding the models to many electrons and near-resonance applications.

\begin{table}\scriptsize
\caption{Effects of geometry and topology in simple graphs. $|\beta|$ is the tensor norm of the first hyperpolarizability and is invariant to orientation of the graphs.  $\beta_{xxx}$ is the value of the first hyperpolarizability's x-diagonal component when the graph is rotated to the position maximizing it.  The minimum value is just the negative of the maximum value.  The range shown for the second hyperpolarizability's x-diagonal component shows the minimum and maximum values, and reveals the expected asymmetry.  Detailed discussions appear in Section \ref{sec:results}.}\label{tab:results}
\newcolumntype{S}{>{\centering\arraybackslash} m{1.3cm} } 
\begin{tabular}{S S c c c c }
  \hline\hline
Graph  & Geometry & Topology & $|\beta|$ & $\left| \beta_{xxx} \right|$ & $\gamma_{xxxx}$ \\
  \hline\hline
\includegraphics{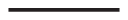} & Line & Line & 0.000 & 0.000 & -0.126 \\
\includegraphics{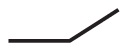} & Bent Wire & Line & 0.172 & 0.172 & -0.126 to 0.007 \\
\hline
\includegraphics{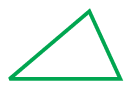} & Triangle & Loop & 0.086 & 0.049 & -0.138 to 0 \\
\includegraphics{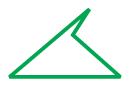} & Simple Quadrangle & Loop & 0.087 & 0.056 & -0.138 to 0 \\
\includegraphics{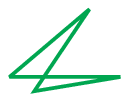} & All Quadrangles  & Loop & 0.087 & 0.069 & -0.138 to 0 \\
\includegraphics{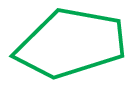} & All Quintangles  & Loop & 0.087 & 0.069 & -0.138 to 0 \\
\hline\hline
\includegraphics{triangle.EPS} & Triangle & Loop & 0.086 & 0.049 & -0.138 to 0 \\
\includegraphics{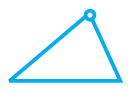} & Triangle & Line & 0.13 & 0.13 & -0.064 to 0.006 \\
\includegraphics{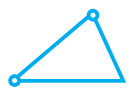} & Triangle & Line & 0.17 & 0.17 & -0.086 to 0.007 \\
\includegraphics{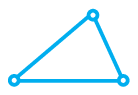} & Triangle & Line & 0 & 0 & -0.114 to 0 \\
\hline
\includegraphics{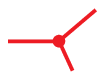} & 3-star & 3-Fork & 0.58 & 0.58 & -0.138 to 0.30 \\
\includegraphics{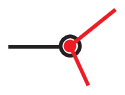} & 3-star & Line & 0.172 & 0.172 & -0.126 to 0.007 \\
\includegraphics{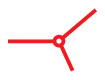} & 3-star & Line & 0 & 0 & -0.126 to 0 \\
\hline\hline
\\
\end{tabular}
\end{table}

\section{Quantum Mechanics of Graphs}\label{sec:NLOgraphs}

\begin{figure}
\includegraphics[width=4.6in]{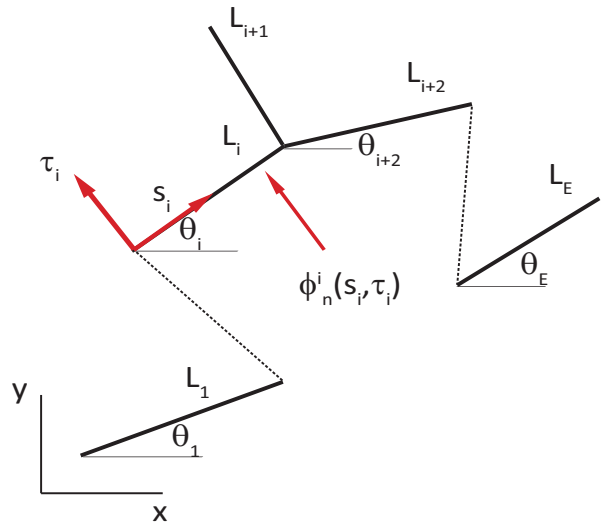}\\
\caption{A multi-edge quantum graph.  Each edge supports eigenfunctions that have the same energy spectrum as the full graph. The eigenfunctions of the graph are unions of the edge eigenfunctions, $\psi_{n}^{i}(s_{i},\tau_{i})$.}\label{fig:graphNEW}
\end{figure}

Quantum graphs are comprised of distinct edges supporting particle dynamics that are governed by a single Hamiltonian.  The graph has a set of energy eigenstates and an associated energy spectrum.  Projections of the eigenvectors onto a particular edge do not form a complete set of eigenfunctions, though they share the same energy spectrum as the full graph.  The wavefucntions along the edges are required to maintain continuity and  conserve flux. Conversely, the union of all the edge wavefunctions represents the full wavefunction of the particle on the graph.  The union operation produces state vectors that span the Hilbert space satisfying completeness, orthonormality, and closure.  This is easily verified by using the Thomas-Reiche-Kuhn (TRK) sum rules \cite{shafe12.01}.  Figure \ref{fig:graphNEW} illustrates the notation used to describe the graph.  The energy eigenstates fully characterize a graph, yielding the energy spectrum, and matrix elements of operators in position space, and depend on the location of its vertices and specification of its edges (connectivity matrix).  The on-edge coordinates $s_{i}$ and $\tau_{i}$ are related to the fixed $x-y$ coordinates by an O(2) transformation through $\theta_{i}$ on that edge.

When specifying a graph, an arbitrary set of axes are selected, and vertices are specified in this frame, called the fixed reference frame. All calculations of transition moments and hyperpolarizabilities are performed in this frame, giving a fixed description of the nonlinear optical response of a graph with a specified geometry (vertices and edge angles) and topology (states and spectra).

The experimentalist is most interested in the largest value of the hyperpolarizability in a lab frame whose $x$-axis is known and usually used to reference the optical field polarizations interacting with the material.  In analogy to the birefringence of a material, we define the preferred alignment of a nonlinear optical graph to be its {\it preferred diagonal} orientation, meaning the graph has been rotated from its initially selected orientation to the one yielding the largest hyperpolarizability.  Using the rotation properties of the tensors, it is straightforward to identify this orientation for any specified graph.  It is typical for $\beta_{xxx}$ and $\gamma_{xxxx}$ to be largest along different axes, so the preferred diagonal orientation of each may be and usually is different.  $\beta_{xxx}$ ($\gamma_{xxxx}$) should be understood as the largest diagonal tensor component of the intrinsic first (second) hyperpolarizability when the graph is in its preferred diagonal frame.  This is often called the molecular frame.  Inorganic crystals have the convenient property that the lab frame and molecular frame coincide.  In the case of organic crystals and dye-doped polymers, the molecules can be arranged in nonaligned fashion -- for example in a herring bone pattern, so that the molecular and lab frames cannot be aligned.  As such, the bulk susceptibilities are determined from a sum over the molecular hyperpolarizabilities with proper O(2) rotation from the molecular frame of each molecule to the lab frame.

Our objective is to calculate the hyperpolarizability tensors for sets of graphs and to study their variations and limits as a function of the geometry and topology of the graphs by establishing a procedure to calculate the transition moments on the graph, and to calculate the hyperpolarizability tensors and manipulate them to find the extreme ranges as shown in table \ref{tab:results}. The tensor analysis enables us to compare and contrast the effects of changing geometry for fixed topology and changing topology for fixed geometry.

The entire process can be summarized as follows: (1)  select a particular kind of graph, specifying the number of vertices and the connecting edges, (2)  generate a random set of vertices, and calculate the lengths of the edges and the angles each makes with the $x$-axis of the graph's coordinate system, (3) solve the Schr\"{o}dinger Equation on each edge of the graph, and (4) match boundary conditions at the vertices and terminal points.  This results in a set of equations for the amplitudes of the wavefunctions on each edge.  The solvability of this set requires that the determinant of the amplitude coefficients vanishes, leading to a secular equation for the eigenvalues.  Since the particle on the graph wanders only on the edges, it has no knowledge of the orientation of the edges; thus, the secular equation and energy spectra depend solely on the lengths of the edges and the type of boundary conditions imposed for the graph.

A detailed discussion regarding the quantum mechanics of simple graphs was presented in \cite{shafe12.01} in terms of a quasi-one dimensional model with a tightly transversely-confined electron moving freely along the edges of a graph. As shown in \cite{shafe11.01}, the residual effects of the vanishing transverse dimension are present in the TRK sum rules but drop out of the hyperpolarizability calculations.  Along any arbitrary edge, labeled by $i$, the electron is described by free-particle states
\begin{equation}\label{edgeWavefunction}
\phi_n^i (s) = A_{n}^{(i)} \sin(k_n s+\eta_{n}^{(i)}) ,
\end{equation}
where $s$ denotes the longitudinal direction, $k_n$ is the wavenumber of the $n^{th}$ quantum state, and $\eta_{n}^{(i)}$ is the accumulated phase from the origin to the $i^{th}$ edge \cite{shafe12.01}. At each vertex, continuity and flux conservation are imposed.  At terminal vertices, we impose Dirichlet boundary conditions, as the physics of quantum graph models for molecular systems should not permit the transmission of flux beyond the extent of the molecule. Eigenstates for the entire graph are formed from unions of the edge wave functions, as described in Ref. \cite{shafe12.01}  in detail.  The mathematics is that of direct sum Hilbert spaces, recently reviewed in Ref. \cite{gnutz10.01} and utilized in \cite{berko06.01}-\cite{dabag02.01}.

The validity of the solution is verified by calculating both longitudinal and transverse contributions to the sum rules, as well as by showing how they contribute to the dipole-free sum-over-states formalism \cite{kuzyk05.02}. The transverse contribution to the first and second hyperpolarizabilities vanish in the limit of tight confinement, so that only longitudinal motion needs to be considered.

Graphs are built by connecting single quantum wires at vertices, creating three basic elements:  Sequential edges, closed loops, and star vertices.  The general method for computing transition moments along sequential edges has been published \cite{lytel12.01}, as has the method for calculating transition moments for closed loops \cite{shafe12.01}.  We show in this section for the first time the transition moments of the star graph and discuss extensions to graphs with all three basic elements.  The result of one such extension, a so-called lollipop graph, is discussed at the end of this section.

One of the most important results of this paper is the delineation of the origin of geometrical and topological effects in quantum graph models of nonlinear optical molecules.  Since we work with metric graphs, each has a length scale which appears linearly in the transition moments and inversely in the wave vectors.  The length scale appears again in the hyperpolarizabilities in a rather complicated way but the hyperpolarizabilities will be normalized to their maximum allowed values so that the results are scale-independent.  As will be shown, the implications are that the graph's length scale drops out of the analysis of intrinsic properties of graphs, enabling a precise delineation of the contributions from the geometric features of the graph (determined by angles) and its topological features (determined by its normalized transition moments and energies).

\subsection{3-Prong Star Graph}

\begin{figure}
  \includegraphics[width=3.4in]{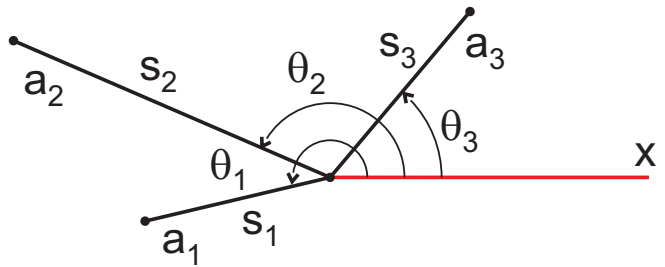}
  \caption{Planar star graph with three prongs.}\label{fig:starGraph}
\end{figure}

The star graph illustrated in Fig. \ref{fig:starGraph} is a simple model of a molecular system but may also comprise an element of a more complex system, such as two rings connected with a prong.  An exact solution for a 3-prong star graph spectrum has been discovered using a periodic orbit expansion \cite{pasto09.01}.  The spectra of star graphs are non-degenerate if at least one ratio of prong lengths is irrational.  When all three prongs are rational numbers, the spectra will contain some doubly degenerate eigenvalues and states.  As previously discussed \cite{pasto09.01}, the rational and irrational cases are physically indistinguishable.  For simplicity, we treat the edges as irrationally-related. At the free ends of each prong, the potential is infinite, and the wavefunction vanishes.

We use the convention of Ref. \cite{shafe12.01}. The energy eigenstates $|N\rangle$ are represented using Dirac notation; but, we use parentheses to represent the vectors, $|N^k)$, which represent the part of the state vector that lives on edge $k$. The eigenstates $|N\rangle$ are orthonormal, but the edge vectors $|N^k)$ are not, or,
\begin{equation}\label{Orthogonality}
\left< M \left| \right. N \right > = \delta_{M,N}  \quad \mbox{and} \quad \left( M^{k} \left| \right. N^{k} \right ) \neq \delta_{M^k N^k}
\end{equation}

The eigenstates $\left | N \right \rangle$ of the star graph may be written as a union of three edge states $|N^{k})$. In coordinate space, we may write
\begin{equation}\label{eigenstates}
\psi_{n}(s)\equiv \langle s|\psi_{n}\rangle = \cup_{i=1}^{3} \left(s_i \right| \left. N \right> = \cup_{i=1}^{3} A_{n}^{(i)}\sin{k_n \left(a_i-s_i \right)}
\end{equation}
where $s_i$ measures the distance from the origin along each edge, $a_i$ are the edge lengths (at least one of which is irrational) and $A_{n}^{(i)}$ are the amplitudes of the wavefunctions in each edge.  `$\cup$' represents the union of the three edges as described above.
At the center node, the wavefunction is continuous, and the net flux in or out of the node is zero.  These conditions yield
\begin{equation}\label{WaveFuncContinuity}
A_n^{(1)}\sin{k_na_1} = A_n^{(2)}\sin{k_na_2} = A_n^{(3)}\sin{k_na_3}
\end{equation}
and
\begin{equation}\label{FluxContinuity}
A_n^{(1)}\cos{k_n a_1}+ A_n^{(2)} \cos{k_n a_2}+ A_n^{(3)} \cos{k_n a_3}=0 .
\end{equation}
For irrational lengths, the arguments of the sinusoidal functions never vanish.  Multiplying the first term in Eq. \ref{FluxContinuity} by $A_n^{(2)} \sin(k_n a_2)A_N^{(3)}\sin(k_Na_3)$, permuting and repeating on the second and third terms, respectively, and using trigonometric addition formulas, the secular equation becomes
\begin{equation}\label{SecularEqn}
\cos{k_n L} = \frac{1}{3}\left[\cos{k_n L_1} + \cos{k_n L_2} + \cos{k_n L_3}\right] ,
\end{equation}
where $L=a_1+a_2+a_3$, $L_1=|a_1-a_2-a_3|$, $L_2=|a_2-a_1-a_3|$, and $L_3=|a_3-a_1-a_2|$.

The solutions to the secular equation for irrational lengths have been discussed at length in Ref. \cite{pasto09.01}, where a periodic orbit expansion was derived for the eigenvalues.  They are nondegenerate and lie one to a cell between root boundaries at multiples of $\pi/L$.  For our purposes, a set of solutions for any finite number of modes is easily found by numerically intersecting the two parts of the secular equation.  In this way, a set of nondegenerate eigenvalues may be obtained for arbitrary (but irrational) prong lengths.

The coefficients $A_n^{(i)}$ of the edge wavefunctions may be found using Eq. (\ref{WaveFuncContinuity}) by recognizing that each term must be independent of the segment index $i$ and equal to the same (wavefunction-dependent) constant, $F_n$.  The eigenfunctions then takes the form
\begin{equation}\label{Eigenfunction}
\psi_n(s) = F_n \cup_{i=1}^3 \frac {\sin{k_n(a_i-s_i)}} {\sin{k_na_i}} .
\end{equation}
Normalizing the total wavefunction given by Eq. (\ref{Eigenfunction}) yields,
\begin{eqnarray}\label{FindF}
1 &=& \int_{graph} ds\psi_n^*(s)\psi_n(s) \\
&=& |F_{n}|^2\sum_{i=1}^3 \frac {\int_0^{a_i}\sin^2(k_n(a_i-s_i))ds_i} {\sin^2(k_na_i)}\nonumber ,
\end{eqnarray}
where this is a sum over the edges of the integrals of $(n^{ k}|s_k \rangle\langle s_k|n^{k})$, which follows from Eq. (\ref{eigenstates}). Performing the integration and using the definition $\sinc(x) = \frac {\sin(x)} {x}$, we get
\begin{equation}\label{TheFs}
|F_n|^{-2} = \sum_{i=1}^3 \frac {a_i} {2} \frac {\left[1-\sinc\left(2k_na_i\right)\right]} {\sin^2{k_na_i}}.
\end{equation}
The amplitudes of the longitudinal wavefunction in each segment may then be expressed as
\begin{equation}\label{Amplitudes}
A_n^{(i)} = \frac{F_n}{\sin(k_na_i)}
\end{equation}
The eigenvalues from Eq. (\ref{SecularEqn}), coupled with the amplitudes from Eq. (\ref{Amplitudes}) and the normalization factors in Eq. (\ref{TheFs}) provide a complete set of eigenstates and spectra for the 3-prong graph with irrational prong lengths.

The dipole matrix elements are calculated via
\begin{eqnarray}\label{xNMgeneral}
x_{nm}=  \langle n|x|m\rangle&=&\left[ \cup_{j=1}^{3} \left(n^j| \right]  x \cup_{i=1}^3 |m^i\right) \nonumber \\
&=& \sum_{i=1}^3 \left(n^i|x|m^i\right).
\end{eqnarray}
Projecting Eq. (\ref{xNMgeneral}) onto longitudinal position space, we may write the individual edge matrix elements as
\begin{equation}\label{xNM}
x_{nm} = \sum_{i=1}^3 A_{n}^{(i)*} A_{m}^{(i)} \cos\theta_iJ^{nm}_i ,
\end{equation}
where $\theta_i$ is the angle between segment $i$ and the x-axis.  The moment integral $J_{nm}^i$ is
\begin{eqnarray}\label{JNM}
J_{nm}^i &=& \int_0^{a_i}s\sin{k_n(a_i-s)}\sin{k_m(a_i-s)}ds \nonumber \\
&=&  \frac {a_i} {2} \left[\sinc^2 \left( \frac {k_{nm}^{-}{a_i}} {2}\right) - \sinc^2\left( \frac {k_{nm}^{+}{a_i}} {2} \right)\right] ,
\end{eqnarray}
where $k_{nm}^{\pm}={k_n}\pm{k_m}$. Eq. (\ref{xNM}) may be numerically evaluated once the eigenvalues are determined using Eq. (\ref{SecularEqn}).  The corresponding $y_{nm}$ may be obtained from Eq. (\ref{xNM}) by replacing the cosine with sine.

We note that the general form of the transition moments for the 3-prong star graph is
\begin{equation}\label{xNMcalc}
x_{nm} = \sum_{i=1}^3 a_i \cos \theta_i K_{nm}^i
\end{equation}
where
\begin{equation}\label{Knm}
K_{nm}^i = K_{nm}^i \left( k_n a_1, k_n a_2, k_n a_3, \frac{a_i} {a_1}, \frac{a_i} {a_2}, \frac{a_i} {a_3}\right).
\end{equation}
Eq. (\ref{xNMcalc}) is expressed in a form that hints at the underlying physics.  If the graph is rescaled so that all lengths are changed by the same factor, $a_i \rightarrow \epsilon a_i$, the wave vectors scale as $k_n \propto \epsilon^{-1}$ and the angles remain the same, so all terms are invariant except for the prefactor $a_i$.  The geometry is defined by the angles, $\theta_i$ and the length ratios, $a_i/a_j$.  Changes in the topology, on the other hand, lead to changes in the boundary conditions, which affect the quantities $k_n a_j$.

This completes the discussion of the general methods for computing states and spectra for graphs that are sequential edges, loops, or stars.  It is instructive to examine exactly how these simple graphs combine to produce a larger graph with its own characteristic function and set of eigenstates, and what additional steps are required when multiple sets of states and spectra are present.

\subsection{Lollipop: A composite graph}

\begin{figure}
\includegraphics[width=3.4in]{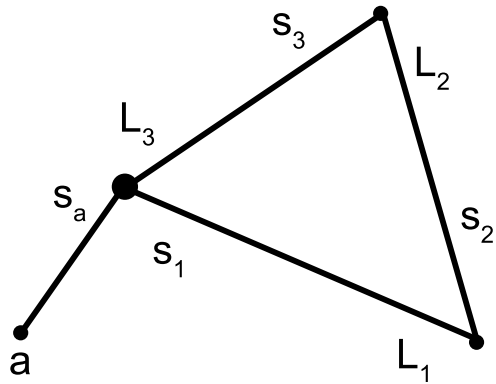}\\
\caption{$N=3$ lollipop graph with a star at one vertex.}\label{fig:1prongLoop}
\end{figure}

\begin{figure}
  \includegraphics[width=3in]{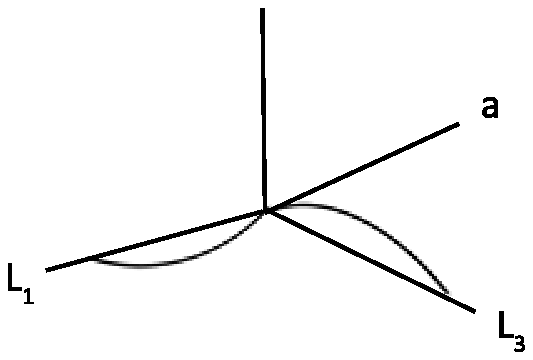}\\
  \includegraphics[width=3in]{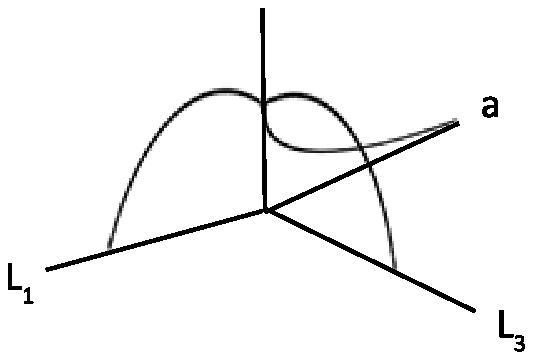}\\
  \caption{Wavefunction when the amplitude is exactly zero on the prong but nonzero in the loop (top) and when the amplitude is nonzero on all edges of the graph (bottom).}\label{fig:1prongStates2}
\end{figure}

Consider a 3-sided loop with a prong at one end, as shown in Fig \ref{fig:1prongLoop}, known as a lollipop.  The edge of length $a$ is assumed to terminate at infinite potential.  The modes of this graph are a composite of two sets of wavefunctions, one set that is nonzero at the central vertex and on all edges, and one for wavefunctions that vanish at the origin and are exactly zero on the prong edge.  Fig. \ref{fig:1prongStates2} shows a representative wavefunction of each type.

Interestingly, the first set correspond to the symmetric wavefunctions of a 3-sided bent wire (open at the central vertex) coupled to a nonzero prong wavefunction, while the second set correspond to the asymmetric wavefunctions of a 3-sided bent wire (open at the central vertex) with a zero prong wavefunction.  Both sets taken together are required to achieve closure, as will be shown by sum rule calculations.  Moreover, each set has a distinct energy spectrum.  For arbitrary values of the prong length $a$ and the total loop length $L=L_1+L_2+L_3$, the energies do not interleave and must be calculated, then ordered from smaller to larger, with the corresponding wavefunctions selected for hyperpolarizability calculations using a finite number of states in the sum-over-states (SOS) expressions.

We begin by writing the eigenfunctions for the graph as a union of edge functions $\psi_n(s)=\cup_{k=1}^4 \phi_n^{k}(s_k)$.  As is usual for sequential edges, we may treat the three-sided closed-loop part of the graph as a wavefunction starting at the central vertex with a continuous amplitude as a function of the distance $L=L_1+L_2+L_3$ along the entire loop \cite{lytel12.01}.  This yields the result
\begin{eqnarray}\label{eigenfunctions}
\psi_n(s) &=& A_n \sin\left[k_n\left(a-s_a\right)\right] \cup \left[ B_{n}\sin\left(k_{n}s_{L}\right) \right.\nonumber \\
&& + \left. C_{n}\cos\left(k_{n}s_{L}\right) \right] ,
\end{eqnarray}
where $s_a$ is measured from the center vertex to the end of the prong definition the stem, the second term in the union represents the union of the three edges in the loop, and $s_L$ is measured from the central vertex along the loop to its return to the central vertex.  Note that the projection of the edges of the loop onto the $x-axis$ changes as $s_L$ moves from one edge to the next.

The boundary conditions yield the two equations
\begin{eqnarray}\label{bc}
\left \{      \begin{array}{ll}
       A_{n}\sin(k_{n}a) = C_{n} = B_{n}\sin(k_{n}L)+C_{n}\cos(k_{n}L) \\
       \\ \\
       -A_{n}\cos(k_{n}a)+B_{n} = B_{n}\cos(k_{n}L)-C_{n}\sin(k_{n}L) .\\
      \end{array}
      \right.
\end{eqnarray}
The solution to these equations provides the amplitudes of the edge functions and the energy eigenvalues of the graph.  As $a\rightarrow 0$, the graph becomes a triangle loop that is open at the central vertex, i. e. a 3-edge bent wire that folds back upon itself.  The eigenfunctions of this limiting case are $G_{n}\sin(k_{n}s_{L})$ with $k_{n}=n\pi/L$, and vanish at both ends of the vertex.

With $A_{n} = 0$, the solution to Eq. (\ref{bc}) corresponds to the even $n$ eigenstates of this `bent wire;' these solutions vanish at the prong with continuous slope through the vertex around the loop, as shown in Fig. \ref{fig:1prongStates2} (top). The odd $n$ solutions of the bent wire are the limiting case of solutions to Eq. (\ref{bc}) that are symmetric on the loop about the central vertex and match to the slope of the nonzero wavefunction on the prong when $A_{n}\neq 0$, as shown in Fig \ref{fig:1prongStates2} (bottom).  Thus, there are two different sets of solutions for the lollipop graph.

The first set is easily found.  Set $A_{n}=0$ in Eq. (\ref{bc}).  This implies $C_{n}=0$ and $k_{n}L=2n\pi$.  The normalized eigenfunctions become
\begin{equation}\label{eigenEven}
\psi_{n}(s_{L})= \left[ \phi_{n}^{(1)}(s_{a}) = 0 \right] \cup \sqrt{\frac{2}{L}}\sin\left(k_{n}s_{L}\right)
\end{equation}
and the energies are $E_{n}=\hbar^{2}k_{n}^2/2m$, with $k_{n}=2n\pi/L$.  The eigenfunctions start at zero value at the center vertex and end at the same value, completing multiples of a full cycle in the loop.

The second set of solutions to Eq. (\ref{bc}) are found by eliminating the three amplitudes to get a secular function given by
\begin{equation}\label{secularL}
f_{sec}=\cos\left[k_n \left(a-\frac{L}{2}\right)\right]-3\cos\left[k_n \left(a+\frac{L}{2}\right)\right].
\end{equation}

Solutions to $f_{sec}=0$  may be found graphically, then interleaved with those of Eq. (\ref{eigenEven}) to produce an ordered set of states and energies for calculating transition moments.  We will not display these moments here, as this graph is not the main focus of this paper.  However, it illustrates the methods required to solve more complex graphs of mixed topology.  It is also an excellent example of how the calculation of the eigenstates and spectrum of a class of graphs may be verified by use of the TRK double-commutator sum rules\cite{shafe12.01}, a powerful method for deciding whether all states and energies have been discovered for a system.

The sum rules along the x-direction may be written as a sum of contributions from the longitudinal part from the s-transition moments along an edge and the transverse part from transition moments perpendicular to the edge along $\tau$ \cite{shafe12.01}:
\begin{eqnarray}\label{sumrules}
&\delta_{\kappa\lambda}& \sum_{i,n}\left[\left( E_n^s - \frac{E_p^s + E_q^s}{2}\right)x_{pn}^{s,i} x_{nq}^{s,i} \right]\nonumber \\
&& + \sum_{i,\nu} \left[\left(E_\nu^\tau - \frac{E_\kappa^\tau + E_\lambda^\tau}{2}\right) I_{pq}^{i} x_{\kappa\nu}^{\tau,i} x_{\nu\lambda}^{\tau,i}\right] \nonumber \\
&=& \frac{\hbar^2}{2m}\delta_{pq}\delta_{\kappa\lambda}.
\end{eqnarray}
The Latin and Greek letters are used to denote longitudinal and transverse components of the wavefunction, respectively. The integral $I_{pq}^{i}$ is the edge state overlap integral in the longitudinal direction and is given by
\begin{equation}\label{Ipq}
I_{pq}^{i} = \int_{i} ds_{i}\phi_{p}^{i}(s_{i})\phi_{q}^{i}(s_{i}).
\end{equation}
Recall that edge functions, $\phi_{p}^{i}(s_{i})$ and $\phi_{q}^{i}(s_{i})$, are not orthogonal on an edge, so Eq. \ref{Ipq} is generally nonvanishing.

Each of the transition moments in Eq. (\ref{sumrules}) contain an angular factor representing the projection of x onto the s-direction $(\cos\theta_{i})$ on edge i and onto the $\tau$-direction $(\sin\theta_{i})$, which is how the geometry of the graph enters into the sum rules. The transition moments depend only on the edge states and energy spectrum of the graph, i. e., the topology.  It is also clear that replacing x with y simply reverses the role of the two terms in Eq. (\ref{sumrules}), with the first term thus equal to the transverse contribution to the sum rules in y-direction and the second term equal to the longitudinal contribution to the y-direction sum rules.  Direct calculation of all quantities not only verifies this, but it is a powerful tool for evaluating the correctness of the states and spectrum.

\begin{figure}
  \includegraphics[width=3.4in]{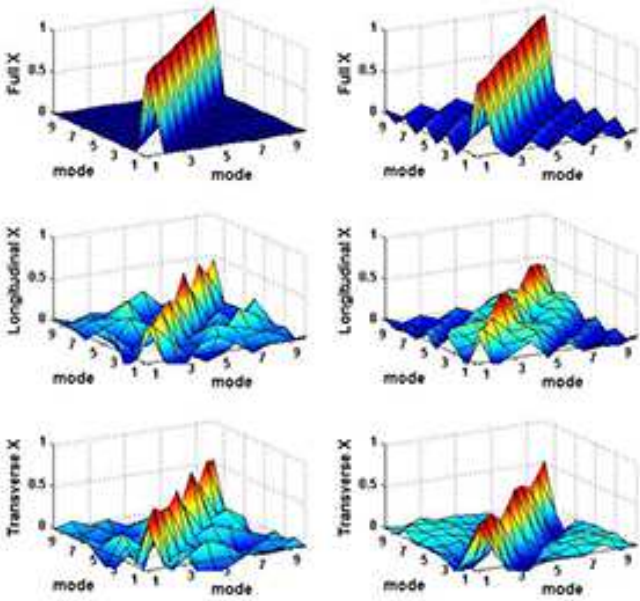}\\
  \caption{(Color online) A plot of the lefthand side of Eq. (\ref{sumrules}) as a function of state number $n$ and $m$.  The left side shows the transverse (bottom), longitudinal (middle), and total sumrule when all states are included.  The right hand side shows the transverse, longitudinal, and total sumrule when an entire set of states is omitted.  The values of the various sums for each pair of modes are indicated on the vertical axes, with an X added to remind the reader that Eq. (\ref{sumrules} was written with x as the longitudinal direction.}\label{fig:sumrule}
\end{figure}

To illustrate the use of sum rules for this purpose, we display in Fig. \ref{fig:sumrule} the results of calculating the lefthand side of Eq. (\ref{sumrules}) for the lollipop graph.  The top left plot shows the total (longitudinal and transverse contributions) sum rules for the graph while the upper right shows the plot with only the single set of edge states computed from Eq. (\ref{bc}) that satisfy Eq. (\ref{secularL}) and represent the states where the particle lives on all edges.  The middle and bottom rows show plots of the longitudinal and transverse contributions to the sum rules.   The missing states are the ones given by Eq. (\ref{eigenEven}), where the particle is excluded from the stem.  The complete set of states exhibit the predicted Kronecker delta behavior.  This exercise illustrates the sensitivity of the sum rules to reveal the absence of even a single low-lying state and is thus a useful tool for validating the correctness/completeness of the energy spectra.

\section{The hyperpolarizability tensors}\label{sec:tensors}

The nonlinear optics of the quantum graph is described by the hyperpolarizability tensors.  Far from resonance, the tenser components of the first hyperpolarizability, $\beta_{ijk}$, a fully symmetric third rank tensor, are given by \cite{orr71.01,kuzyk05.02}
\begin{equation}\label{sh-betaijk}
\beta_{ijk} = -\frac{e^3}{2} P_{ijk} {\sum_{n,m}}' \frac{1}{E_{n0} E_{m0}} r_{0n}^i \bar{r}_{nm}^j r_{m0}^k
\end{equation}
where the prime indicates that the ground state is excluded from the summation, the superscripts $i$, $j$ and $k$ can take on $x$, $y$ or $z$ -- the Cartesian components, $P_{ijk}$ permutes all the indices in the expression, $\bar{r}_{nm} = r_{nm} - r_{00}\delta_{nm}$, and $E_{nm} = E_n - E_m$ is the energy between two eigenstates $n$ and $m$.

Similarly, $\gamma_{ijkl}$ is a fully symmetric, fourth rank tensor and has five independent Cartesian components \cite{bance10.01}. $\gamma_{ijkl}$ is given by \cite{orr71.01}
\begin{eqnarray}\label{gammaijkl}
\gamma_{ijkl} &=&  \frac{1}{6} P_{ijkl}  \left({\sum_{n,m,l}}' \frac{r_{0n}^{i} \bar{r}_{nm}^{j}\bar{r}_{ml}^{k}r_{l0}^{l}}{E_{n0} E_{m0} E_{l0}} \right. \nonumber  \\
&& \left. - {\sum_{n,m}}' \frac{r_{0n}^{i}r_{n0}^{j}r_{0m}^{k}r_{m0}^{l}}{E_{n0}^2 E_{m0}}\right),
\end{eqnarray}
where the permutation operator here is over the four indices $(i,j,k,l)$.  For both of these expressions, the matrix elements $r_{nm}$ are computed in an appropriate orthonormal basis set for the graph, with the integrations taken along the path of the graph with $r(s)$  defined by the sum of the projections of $s$ onto the $x$-axis $(r=x)$ or the $y$-axis $(r=y)$.

The intrinsic values of the first and second hyperpolarizabilities are defined as the ratio of $\beta_{ijk}$ and $\gamma_{ijkl}$ over their fundamental limits, $\beta_{max}$, and $\gamma_{max}$, respectively, or,
\begin{equation}\label{IntrinsicBetaGamma}
\gamma_{ijkl} \rightarrow \frac {\gamma_{ijkl}} {\gamma_{max}} \hspace{2em} \beta_{ijkl} \rightarrow \frac {\beta_{ijkl}} {\beta_{max}} .
\end{equation}
The fundamental limits are the highest attainable first and second hyperpolarizabilities and solely depend on the number of electrons, $N$, and the energy gap between the ground and the first excited state, $E_{10}$. They are given by \cite{kuzyk00.01, kuzyk00.02}
\begin{equation}\label{sh-betaMax}
\beta_{max} = 3^{1/4} \left(\frac{e\hbar}{m^{1/2}}\right)^3 \frac{N^{3/2}}{E_{10}^{7/2}}
\end{equation}
and
\begin{equation}\label{sh-gammaMax}
\gamma_{max} = 4 \left(\frac{e^4\hbar^4}{m^2}\right) \frac{N^{2}}{E_{10}^{5}} .
\end{equation}

The intrinsic hyperpolarizabilities have the property that they are the same for all graphs of the same shape, $i.e.$ for all graphs whose edge lengths are rescaled according to $a_i \rightarrow \epsilon a_i$.  Artifacts due to simple size effects are eliminated by using these intensive quantities.  As such, {\bf all first and second hyperpolarizabilities discussed beyond this point are implicitly the intrinsic values} unless specifically stated otherwise.

We can identify the independent components for $\beta$ as the set $(xxx,xxy,xyy,yyy)$ and for $\gamma$ as the set $(xxxx,xxxy,xxyy,xyyy,yyyy)$. The components are measured in some specified reference frame and are related to those in another frame by a suitable rotations of the hyperpolarizability tensors.  The fully symmetric $\beta$ tensor requires knowledge of four components in one frame to know them all in any other frame, while the fully symmetric fourth-rank tensor $\gamma_{ijkl}$ requires five components.  The determination of the nonlinear optical properties of the graph is thus reduced to the calculation of the tensors for graphs with a specific geometry for which states and spectra are known.

We saw in Eq. (\ref{xNMcalc}) that the transition moments take the form
\begin{equation}\label{xNM2}
x_{nm} = \sum_{i} a_{i}\cos\theta_{i} K_{nm}^{i} , \\
\end{equation}
where $i$ is summed over all edges of the graph and $\theta_{i}$ is the angle between the segment $i$ and the external $x$-axis.  A similar expression holds for $y_{nm}$ with $\cos\theta_{i} \rightarrow \sin\theta_{i}$.  The dimensionless factor $K_{nm}^{i}$ for star graphs contains an integral of edge wavefunctions with distance along the edge and depends on dimensionless products of the form $k_{n}a_{i}$.  This delineation holds for every quantum graph with the caveat that the explicit expression for the $K_{nm}^{i}$ takes on forms specific to the topology of the graph.  The number of eigenfunctions, their degeneracies (if any), and the energy spectra are fixed by the boundary conditions imposed by the topology of the graph.  The angular factors in Eq. (\ref{xNM2}) describe the geometry of the graph and are identical for graphs with identical geometries but different topologies, such as a closed-loop triangle graph and a triangle graph with identical shape but one open vertex.

It is now clear how analyzing classes of graphs with similar geometry but different topology, and vice-versa, enables extraction of both the topological and geometrical effects on the nonlinear optics of the graph.  For example, the diagonal component of the first hyperpolarizability tensor will always take the form
\begin{eqnarray}\label{betaGeoTop}
\beta_{xxx} = \sum_{i,j,k} \left(\cos\theta_{i} \cos\theta_{j} \cos\theta_{k}\right)\times A_{ijk},
\end{eqnarray}
where
\begin{eqnarray}
A_{ijk} \sim \sum_{n,m}' \frac{K_{0n}^{i} \bar{K}_{nm}^{j} K_{m0}^{k}}{E_n E_m}.
\end{eqnarray}

Equation (\ref{betaGeoTop}) expresses the influence of the angular factors describing the geometry of the graph on each segment's contribution to the underlying quantum mechanics of the full graph as embodied in the edge factors $K_{nm}^{i}$.  It is thus reasonable to speak of the cosine factors as the geometric specifiers.  Since the topology of the graph determines the boundary conditions on the eigenstates, topological effects originate solely in $A_{ijk}$.

$A_{ijk}$ has three indices that couple the angular factors in ways that are determined by the topology of the graph, not its geometry.  This means that the differences between a closed loop triangle and one with an open vertex arises from the quantum states, i.e. the topology of the graph.  Along the edges of the graph, the electron knows nothing about angles; it only knows about the value of its wavefunction on any particular edge.

As noted above, the properties of a graph are intrinsic to it and do not depend upon the lab coordinate system.  However, the measured values in the lab are related to those on the graph by the rotation group.  For planar graphs, we may restrict our focus to O(2) rotations.  A calculation in the frame of the graph using $x_{nm}$ and $y_{nm}$ will yield the hyperpolarizability tensors in that frame, but referred to the x-axis in the graph's frame.  It is likely that this axis will not be the one for which any particular tensor component is maximized.  In fact, there is no way to know how to pick the x-axis to yield the best x-component of the tensor, for example. This is irrelevant, of course, because the tensor in one frame is related to that in another by its transformation properties under O(2).  This same remark holds for the relationship between the tensors in the graph's frame and their values in the lab frame.  From this discussion, it is seen that the tensor properties under O(2) transformations may be used to determine the effect of the graph's geometry on the hyperpolarizability tensors when it is calculated in one frame and measured in another.

Assuming that the independent tensor components of the first and second hyperpolarizabilities in any arbitrary frame of coordinates are given, they can be used to calculate the tensor components in any other reference frame. Specifically, for a reference frame that is rotated $\phi$ degrees with respect to the initial reference frame, the diagonal components, $\beta_{xxx}(\phi)$ and $\gamma_{xxxx}(\phi)$, can be determined using
\begin{eqnarray}\label{BetaCartesian}
\beta_{xxx}(\phi) &=& \beta_{xxx}\cos^3\phi \nonumber + 3\beta_{xxy}\cos^2\phi \sin\phi \nonumber \\ &+&  3\beta_{xyy}\cos\phi \sin^2\phi + \beta_{yyy}\sin^3\phi,
\end{eqnarray}
and
\begin{eqnarray}\label{GammaCartesian}
\gamma_{xxxx}(\phi) &=& \gamma_{xxxx}\cos^4\phi+4\gamma_{xxxy}\cos^3\phi\sin\phi \nonumber \\ &+& 6\gamma_{xxyy}\cos^2\phi \sin^2\phi + 4\gamma_{xyyy}\cos\phi \sin^3\phi \nonumber \\
&+& \gamma_{yyyy}\sin^4\phi , \nonumber
\end{eqnarray}
where $\beta_{xxx}$ ($\gamma_{xxxx}$) is by definition at an extreme value when the graph is rotated through $\phi$.  Once the graph is solved and the tensor components are known in its frame, $\phi$ is easily found by maximizing Eq. (\ref{BetaCartesian}) for $\beta_{xxx}$ and Eq. (\ref{GammaCartesian}) for $\gamma_{xxxx}$.

The most important quantity, however, are tensor norms. The tensor norms are invariant under any transformation and provide immediate insight into the limiting responses of the graphs.  They are given by,
\begin{equation}\label{BetaNorm}
|\beta| = \left( \beta_{xxx}^2+3\beta_{xxy}^2+ 3\beta_{xyy}^2+\beta_{yyy}^2\right)^{1/2}
\end{equation}
and
\begin{equation}\label{GammaNorm}
|\gamma| = \left( \gamma_{xxxx}^2+4\gamma_{xxxy}^2+ 6\gamma_{xxyy}^2+4\gamma_{xyyy}^2+\gamma_{yyyy}^2\right)^{1/2}
\end{equation}
These are the magnitudes of the graph's hyperpolarizabilities and are both scale and orientation-independent.

The use of tensors to extract the nonlinear optical response as a function of geometry and topology is most easily achieved by transforming the Cartesian tensors to spherical tensors.  Zyss \emph{et al} \cite{zyss94.01} and Joffre \emph{et al} \cite{joffr92.01} discuss the molecular nonlinearities in multipolar media using irreducible, spherical representations for $\beta$, an approach which provides insight into the shape-dependence of the first hyperpolarizability, particularly with respect to certain symmetry groups. The transformation from a Cartesian to a spherical tensor representation is achieved using Clebsch-Gordon coefficients and has been extensively discussed in the literature \cite{jerph78.01}.

For fully symmetric Cartesian tensors, $\beta$ has a vector $(J=1)$ and a septor $(J=3)$ component \cite{jerph78.01}.  Its irreducible representation is $1 \bigoplus 3$, with a total of four independent Cartesian components, as noted earlier.  Similarly, $\gamma$ has a scalar, deviator, and nonor component, and its irreducible representation is $0 \bigoplus 2 \bigoplus 4$.  The specific form of the spherical tensor expansions may be directly calculated using the Clebsch-Gordon coefficients.  Using the method described in Ref. \cite{jerph78.01}, we get
\begin{eqnarray}\label{BetaTensors}
S_{\pm1}^{1} &=& \sqrt{(3/10)\left[\pm \left(\beta_{xxx}+\beta_{xyy}\right) +\imath \left(\beta_{yyy}+\beta_{xxy} \right)\right]} \nonumber \\
S_{\pm3}^{1} &=& \sqrt{(3/40)\left[\pm \left(\beta_{xxx}+\beta_{xyy}\right) +\imath \left(\beta_{yyy}+\beta_{xxy}\right)\right]} \\
S_{\pm3}^{3} &=& \sqrt{(1/8)\left[\pm(-\beta_{xxx}+3\beta_{xyy}) +\imath(\beta_{yyy}-3\beta_{xxy})\right]}.\nonumber
\end{eqnarray}

Similarly, the spherical components of $\gamma$ are given by
\begin{eqnarray}\label{GammaTensors}
T_{0}^{0} &=& \sqrt{(1/5)\left[\gamma_{xxxx}+2\gamma_{xxyy}+\gamma_{yyyy}\right]} \nonumber \\
T_{2}^{0} &=& \sqrt{(1/7)\left[\gamma_{xxxx}+2\gamma_{xxyy}+\gamma_{yyyy}\right]} \nonumber \\
T_{\pm2}^{2} &=& \sqrt{(3/14)\left[(-\gamma_{xxxx}+\gamma_{yyyy}) \mp2\imath(\gamma_{xxxy}+\gamma_{xyyy})\right]} \nonumber \\
T_{0}^{4} &=& \sqrt{(9/280)\left[\gamma_{xxxx}+2\gamma_{xxyy}+\gamma_{yyyy}\right]} \nonumber \\
T_{\pm2}^{4} &=& \sqrt{(1/28)\left[(-\gamma_{xxxx}+\gamma_{yyyy}) \mp2\imath(\gamma_{xxxy}+\gamma_{xyyy})\right]} \nonumber \\
T_{\pm4}^{4} &=& \Big( \left(1/4 \right) \left[ \left(\gamma_{xxxx}-6\gamma_{xxyy}+\gamma_{yyyy}\right) \right. \nonumber \\
&& \left.  \pm 4 \imath \left(\gamma_{xxxy}-\gamma_{xyyy}\right)\right] \Big)^{1/2}
\end{eqnarray}
where $\imath^2 = -1$.

The total tensor norm for $\beta$ is found by summing $|S_{m}^{J}|^2$ over the $2m+1$ components $m=-J,-J+1,...0...J-1,J$ for $J=1$ and $J=3$.  Similarly, the total tensor norm for $\gamma$ is found by summing $|T_{m}^{J}|^2$ over the $2m+1$ components $m=-J,-J+1,...0...J-1,J$ for $J=0$, $J=2$, and $J=4$.  These norms are of course identical to those that would be computed from the original Cartesian tensors. But the new information here is that we now have a geometric description of the rotation properties of graphs as a function of their shape that can display their most significant contributions in terms familiar to designers of nonlinear optical molecules.  The significance of this will be displayed in the next section.

\section{Geometric and topological effects}\label{sec:results}

Each of the graphs in Table \ref{tab:results} were analyzed for their nonlinear optical properties using the eigenstates and energy spectra described in past sections for each topology with the tensor formalism described in Section \ref{sec:tensors}.  For every graph topology, a set of eigenstates and energies was calculated for tens of thousands of random geometries, and the hyperpolarizabilities were calculated for each sample.  These methods allowed extraction of the extreme values shown in Table \ref{tab:results}.  The tensors allow extraction of the type of component that most contributes to a graph, e. g., vector, 3-tensor, etc. as well as permitting a direct comparison among fixed topologies with different geometries and a cross-comparison of identical geometries across different topologies.  The results are discussed next.

\subsection{Bent wire topologies}

The simplest graph in the first topologically equivalent set is the line segment shown in the first row of Table \ref{tab:results}, and is characterized by vanishing $\beta_{xxx}$ by virtue of its centrosymmetry, and by $\gamma_{xxxx} = -0.126$.  A bent wire, as shown in the second row of Table \ref{tab:results}, is topologically equivalent to the line but of different geometry (i.e. shape).  The largest hyperpolarizability found is given by $\left| \beta_{xxx} \right| = 0.172 $. The second hyperpolarizability varies from -0.126 to +0.007.  This shows how the hyperpolarizability smoothly varies from zero for the straight wire and peaks for a bent wire that turns back on itself.  For the case of $\gamma_{xxxx}$, the best shape is given by an acute angle with vertex $(0,0)$ and endpoints $(1,-1)$ and $(1.8,0)$.  Despite their simplicity, bent wires have an intrinsic first hyperpolarizability that is about $1/6^{th}$ of the fundamental limit.  In contrast, most of the best planar molecules fall a factor of 30 short of the limit.\cite{kuzyk03.02}

Adding a third segment to a bent wire with no constraint on its orientation yields about the same extremes of $\beta_{xxx}$ and $\gamma_{xxxx}$ as the two-wire graph.  Imposing geometrical constraints on the third edge causes minor quantitative change, as can be seen in the line triangles with open vertices in Table \ref{tab:results}. The addition of even more degrees of freedom enables wire configurations of optimum shape to be marginally improved and provides limited enhancement of nonlinearities.  For bent wires, $\gamma_{xxxx}$ can be positive or negative.  The eigenstates of bent wires are non-degenerate, a consequence of the open topology of the graph.

\subsection{Closed loop topologies}

Consider next the closed-loop topologies from Table \ref{tab:results}. Owing to the periodic boundary conditions of a loop, the solutions are a doubly-degenerate set of eigenstates with a non-degenerate ground state of zero energy. (Note that a complex quadrilateral has the same topology as a simple quadrilateral when the crossing edges do not transfer probability flux, as we assume here).

The closed-loop topology severely limits the magnitudes of the first hyperpolarizability across the set of geometries.  Compared to a simple bent wire, a triangle loop is a poorer nonlinear optical structure with a maximum $\beta_{xxx} \sim0.049$.  More significant, the intrinsic second hyperpolarizability $\gamma_{xxxx} $ is always less than or equal to zero in closed-loop graphs.  Finally, rapid saturation of the nonlinearity with the number of edges occurs, as it did for bent wires. Large $\beta_{xxx} $ is associated with open, isosceles-like shapes, while low-aspect ratio (flat) triangles yield zero $\beta_{xxx} $.  But $\gamma_{xxxx} $ is the most negative for flat triangles.  Quadrangles and above may be geometrically squeezed into these shapes, thus explaining how the topological features of the loops drive most of the physics, while the geometric shapes have only a modest effect.

\begin{figure}
\includegraphics{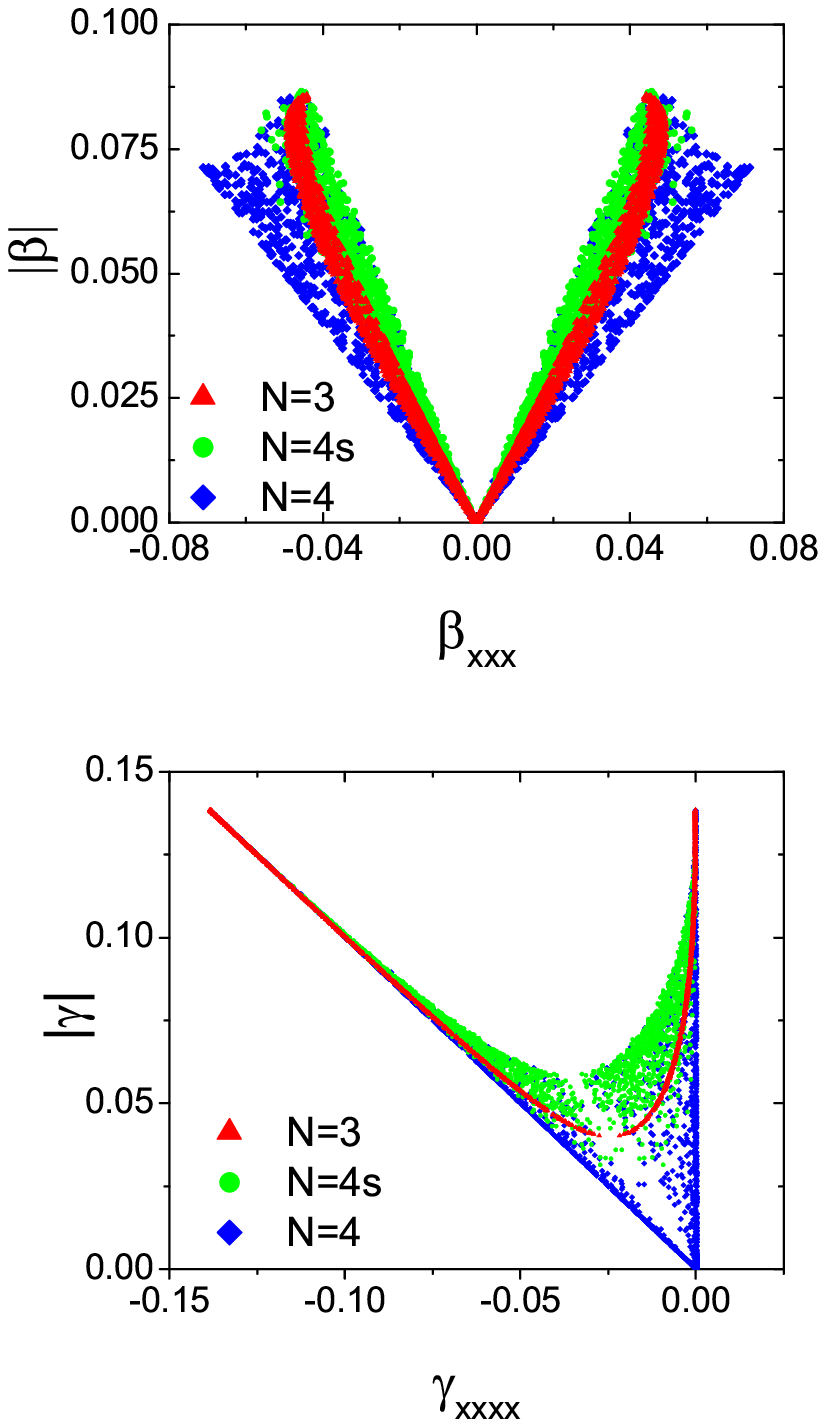}
\caption{(Color online) $| \beta |$ vs $\beta_{xxx}$ and $\left| \gamma \right|$ vs $\gamma_{xxx}$ for 10,000 randomly-sampled graph configurations of fixed topology.  Quintangle results are not shown, as their tensors look almost identical to the quadrangles.}
\label{fig:loopTensorNorms}
\end{figure}

Figure \ref{fig:loopTensorNorms} compares the tensor properties of the distribution of 10,000 random samples of configurations of the closed-loop triangle ($N=3$), simple quadrangle ($N=4s$), and complex quadrangle ($N=4$) graphs shown in Table \ref{tab:results}.  (Note that all plots that follow use 10,000 random samples.)  The plot on the top shows $| \beta |$ vs $\beta_{xxx}$, while the plot on the bottom shows $| \gamma |$ vs $\gamma_{xxxx}$ for the three (fixed topology) geometrically-different closed loops.  Triangle graphs, which have only two angular degrees of freedom, have large (absolute) projections onto the x-axis only when $| \beta |$ is large as well.  The quadrilateral, having 3 degrees of freedom, has the largest value of $\beta_{xxx} $, but marginally so.  When $| \beta |$ has its largest value,  $\beta_{xxx} $ is not optimal.  However, $| \beta |$ is maximum for the same values of $\beta_{xxx} $ in all geometries.

The constraint imposed by the closed triangular loop on the second hyperpolarizability yields a tight grouping of all configurations, as shown by the $N=3$ plot.  Adding extra degrees of freedom yields a greater spread in $| \gamma |$ for a given value of $\gamma_{xxxx}$, as is also found for the first hyperpolarizability.  However, $\gamma_{xxxx}$ is at its minimum and maximum when $| \gamma |$ is at its maximum.  When $| \gamma |$ is minimum, $\gamma_{xxxx}$ appears to be minimum only in the limit of an infinite number of degrees of freedom.

In summary, the range of $\beta$ and $\gamma$ over all configurations is limited by the loop topology, not the geometry.

\begin{figure}
\includegraphics{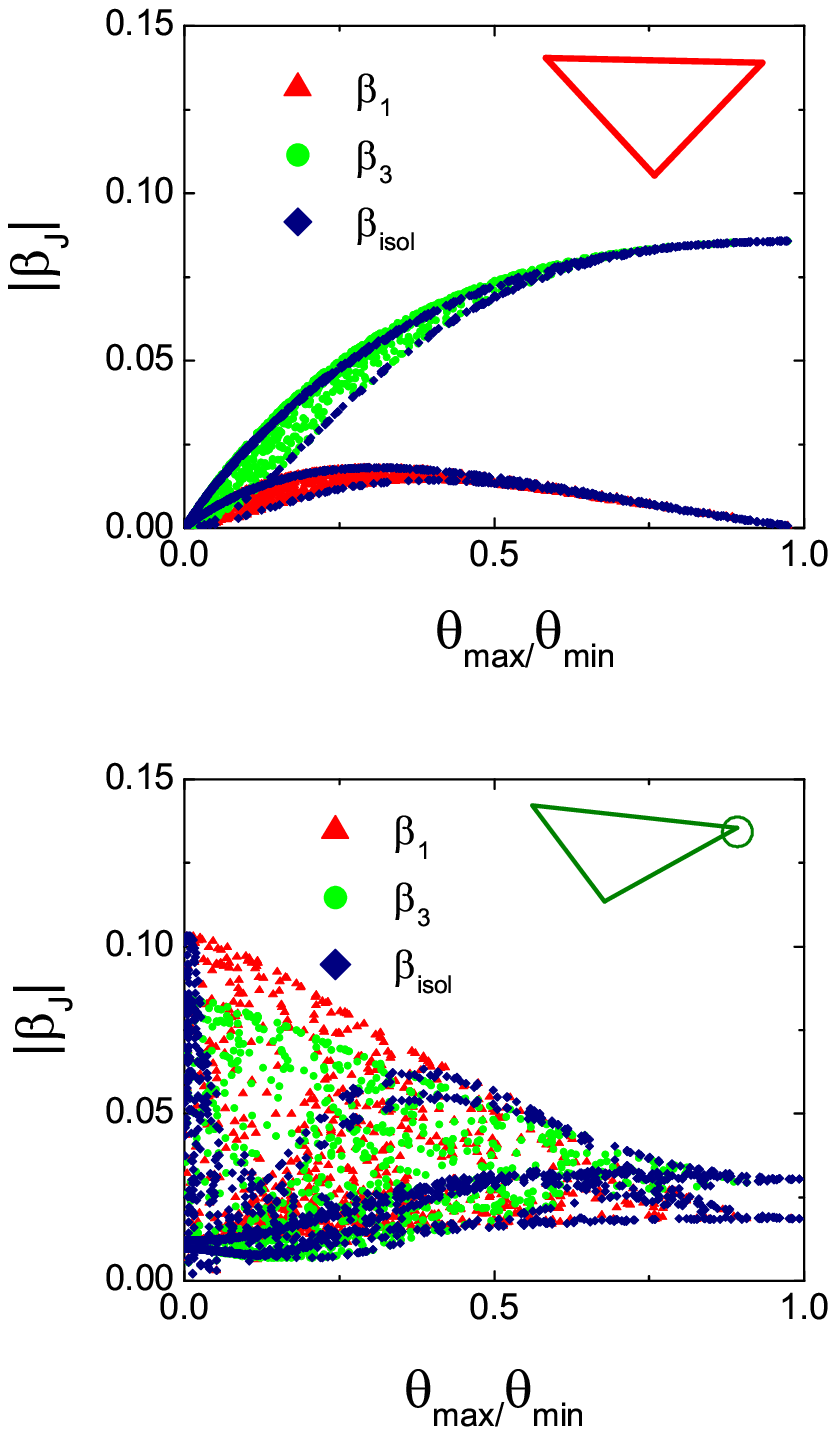}
\caption{(Color online) Spherical tensors for $| \beta |$ as a function of the ratio of the smallest angle $\theta_{min}$ to the largest angle $\theta_{max}$ in a closed loop (top) and open (bottom) triangle graph.}
\end{figure}

\subsection{Closed and open topologies in loop graphs}

\begin{figure}
\includegraphics[width=3.4in]{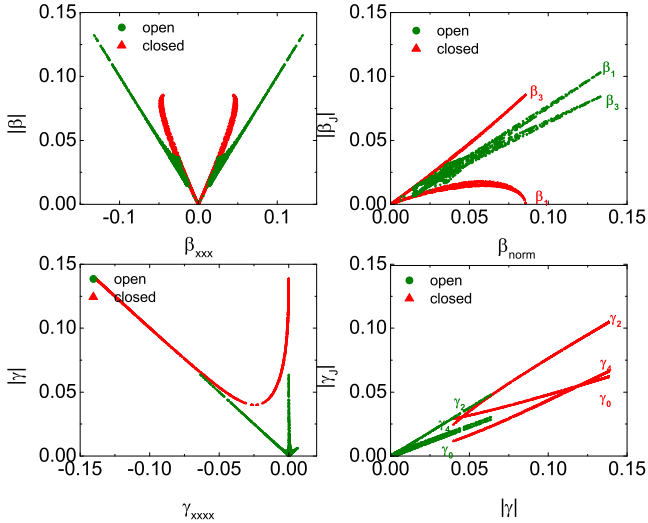}
\caption{(Color online) Topological dependence for fixed geometry.  Shown are the variation in the hyperpolarizability tensor components and their norms for open and closed triangle graphs.}
\label{fig:triTensorsOpenClosed}
\end{figure}

Figure \ref{fig:triTensorsOpenClosed} shows a plot of the contributions to $|\beta|$ from the two spherical tensors for a closed triangle and an open vertex triangle. For closed loops, large $| \beta |$ graphs are dominated by the $\beta_3$ tensor when $\theta_{min} = \theta_{max}$ implying that $\beta$ is maximal for an equilateral triangle.  The dark blue points are for isosceles triangles.  The conclusion is that high aspect ratio triangles have near-zero $\beta_{xxx}$ regardless of their orientation.

For triangles with an open vertex, corresponding to a change in topology from the closed triangle, the relationship between the spherical tensor components and the angle ratio are profoundly different.  $\beta_1$ exceeds $\beta_3$ and $| \beta |$ is substantially larger for the open vertex triangles.  Furthermore, isosceles triangles no longer give the largest $\beta$ and the equilateral triangle has smallest $| \beta |$, compared with largest $ | \beta | $ in the case of the closed triangle.  Topology therefore has a profound effect on both the magnitude and character of the first hyperpolarizability.

The effects of topology on the first and second hyperpolarizability are emphasized in Figure \ref{fig:triTensorsOpenClosed}.  For the closed loop, both $\beta_{xxx}$ and $| \beta |$ are smaller; and, the range of $\beta_{xxx}$ is substantially less than for $| \beta |$.  When a vertex is opened,  $\beta_{xxx}$ and $| \beta |$ are both larger, and there exists a configuration for which $\beta_{xxx} =| \beta |$ for all $| \beta |$.  The opposite is true for $\gamma$.  The closed configuration yields the largest value of $| \gamma |$ and $\gamma_{xxxx}$.  Additionally, the same span of geometrical configurations for each leads to divergent behavior as shown in the figures.

$\beta_1$ is always smaller than $\beta_3$ in the closed triangle but are both comparable in the open one.  The second hyperpolarizability $\gamma$ is dominated by $\gamma_2$ in all cases, and each $\gamma_i$ in the loop topology approximately parallels $\gamma_i$ in the open vertex case; but, the values of $\gamma_i$ are larger for the loops.

These observations can be understood as follows.  Opening a single vertex in a closed triangle removes half the eigenstates and shifts the ground state to nonzero energy.  The result is a shift in the vector component of $| \beta |$ from an insignificant contributor in closed triangles to a substantial contributor for the open triangles.  The topological change causes two significant changes in $\gamma_{xxxx}$, as it allows for configurations with exactly zero norm as well as shapes with positive $\gamma_{xxxx}$ for open triangles.  Though geometrically identical, the closed and open topologies differ by over a factor of three in $\beta_{xxx}$, and the open triangles have configurations with positive $\gamma_{xxxx}$.

For simple quadrangles, the distribution of $\beta$ fills a greater area in the $|\beta_J|-|\beta|$ plane than for triangles, reflecting the additional degree of freedom in its configuration space.  In a triangle, two angles define a set of similar triangles -- all having the same geometry and same intrinsic nonlinearities.  In a quadrangle, three angles do not form a unique similarity class, leading to a greater spread of intrinsic nonlinearity.  The additional degrees of freedom afforded to a general quadrangle fills even a greater part of the plane, yet $\beta_{xxx}$ increases only modestly with increasing degrees of freedom.  The loop topology constrains $\beta_{xxx}$ to values well under that of the simple bent wires and keeps $\gamma_{xxxx}$ negative for any shape.

Table \ref{tab:results} summarizes these results and compares the closed-loop topology with three line configurations, each of which is a bent wire or a composite of bent wires discussed previously.  Opening a single vertex in a closed-loop triangle eliminates the eigenstate degeneracies and produces a ground-state with nonzero energy.  The bent wire triangle with one open vertex and no flux circulation has a much larger $\beta_{xxx} \sim0.133$, and a positive, maximum $\gamma_{xxxx} $.  In these triangle graphs, the extremes of $\beta_{xxx}$ and $\gamma_{xxxx}$ occur for the opposite geometries of those in the closed-loop triangle.

\begin{figure}
\includegraphics{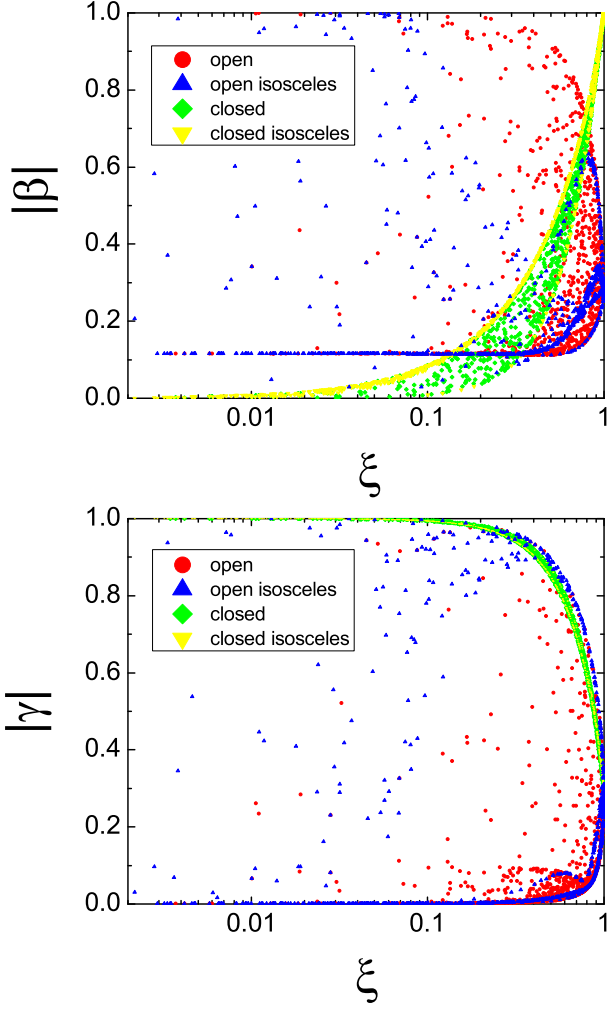}
\caption{(Color online) Scaling of $| \beta |$ (top) and $| \gamma |$ (bottom), both normalized to unity, as a function of geometric parameter $\xi$, for closed- and open-loop topologies.  The blue points represent isosceles-like geometries.}
\label{fig:TriAreaNormCompare}
\end{figure}

To quantify the geometrical effect of the {\it openness} of a loop, we define the dimensionless geometrical parameter $\xi$ as the ratio of the area of a loop to the square of its perimeter.  Since a polygon with all edge lengths equal to each other and all angles equal to each other has maximum $\xi$, we normalize all $\xi$ to this value to get the intrinsic geometric parameter.  Figure \ref{fig:TriAreaNormCompare} shows $| \beta |$ and $| \gamma |$ (normalized to unity) as a function of $\xi$ for both closed (loop) and open (bent wire, one vertex open) triangles.  The open circles are isosceles-like triangles and $\xi=1$ is an equilateral triangle. Opening a second vertex in the bent-wire triangle increases the $\beta_{xxx}$ toward that of a bent wire while still yielding positive $\gamma_{xxxx}$ configurations. Opening all vertices produces three single wires, for which $\beta_{xxx}$ is exactly zero and $\gamma_{xxxx}$ approaches the value it has for a single bent wire.  In such systems, the eigenstates are constructed from three 1-wire graphs by ordering the individual energies of each wire to form the composite and demanding that superpositions maintain unitarity.

\subsection{Closed and open topologies in star graphs}

\begin{figure}
\includegraphics{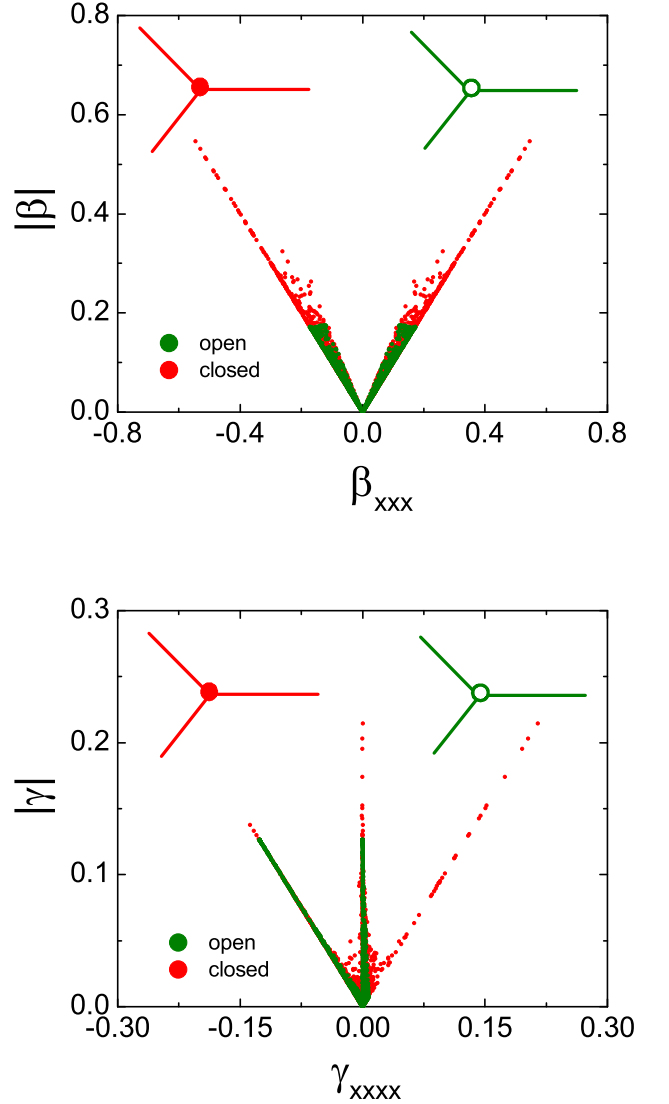}
\caption{(Color online) Topological dependence for fixed geometry of star graphs, showing on the left (right) a very large enhancement in $\beta_{xxx}$ ($\gamma_{xxxx}$) for the closed star with three attached prongs (red) compared to the same graph with one prong detached (green).}
\label{fig:starPanel}
\end{figure}

The final set of graphs in Table \ref{tab:results} are the star graphs and are fundamentally different than any discussed so far. The 3-prong star graph is an irregular graph with four degrees of freedom (two angles and the lengths of two edges), providing much larger nonlinearities than any of the regular graphs, with a maximum $\beta_{xxx} $ of order $0.57$, over half of the fundamental limit and the largest calculated to date for any simple structure.  Figure \ref{fig:starPanel} illustrates the dramatic change caused by a simple topological shift in a graph from three connected prongs to two.  As expected, a major change in the topology of the star graph by opening a prong at the central vertex changes the highly-active 3-fork into a graph resembling a bent wire system.  An obvious conclusion of this analysis is that combinations of simple structures such as bent wires and stars may lead to structures with even larger nonlinear optical response.  In fact, the lollipop graph has a maximum $\beta_{xxx} $ of order $0.61$, larger than the basic star. More complex structures are under study.

\subsection{Universal Properties}

We conclude this section with a brief discussion of the results using a potential energy model \cite{zhou06.01,zhou07.02, ather12.01}. Studies which optimize the shape of the potential energy function yield the largest possible nonlinear optical response of $\beta_{xxx}  \simeq 0.71$ for a large set of potentials, which universally have the property that $X = x_{01}/x_{01}^{max} \simeq 0.79$ where $x_{01}^{max}$ is given by
\begin{equation}\label{Xmax}
x_{01}^{max} = \left(\frac{\hbar^2}{2 m E_{10}}\right)^{1/2} ,
\end{equation}

In the three-level ansatz, the normalized first hyperpolarizability $\beta_{int}$ can be expressed as \cite{kuzyk09.01}
\begin{equation}\label{betaIntfG}
\beta_{int} = f(E)G(X) ,
\end{equation}
where
\begin{equation}\label{f(E)}
f(E) = (1-E)^{3/2} \left( E^2 + \frac {3} {2} E + 1 \right),
\end{equation}
and
\begin{equation}\label{G(X)}
G(X) = \sqrt[4]{3} X \sqrt{\frac {3} {2} \left( 1 - X^4\right)},
\end{equation}
and $E=E_{10}/E_{20}$.

\begin{figure}
\includegraphics[width=90mm]{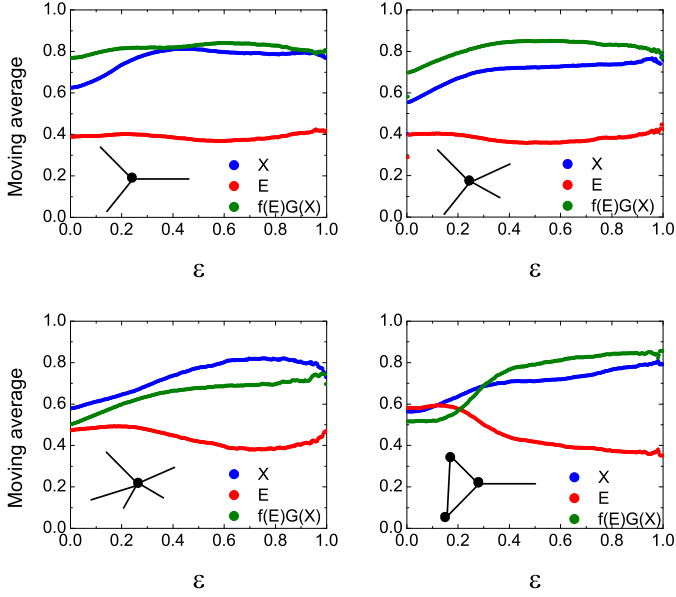}
\caption{(Color online) Scaling of X and E as $\beta$ increases toward its maximum value for three star graphs and a lollipop with a star vertex.  The moving average is over values having $\beta_{xxx}$ greater than $\epsilon$ times its maximum value on the specific graph.  Despite their geometric differences, all four graphs asymptote to universal values, indicating that the topology of the star vertex dominates the spectrum of each graph, including the lollipop (despite its loop).}
\label{fig:XandE}
\end{figure}

This expression is valid as the hyperpolarizability approaches its maximum intrinsic value.  But it also appears to apply universally across different star graphs, as shown in Fig. \ref{fig:XandE}.  This would seem to imply that graphs with the highest nonlinearities have near-optimum potentials.  A detailed study of the contributing states near the maxima of each graph topology verifies that the three level ansatz is valid.

Another result seen from Fig. \ref{fig:XandE} is that the lollipop graph behaves like a star graph (and $\beta_{xxx}  \simeq 0.61$), not a loop (for which E is constant and $\beta_{xxx}  \simeq 0.05$). The star graph topology sets the global behavior of graphs containing it, suggesting that ring molecules with prongs may exhibit much larger hyperpolarizability than rings by themselves.

\begin{figure}
\includegraphics[width=90mm]{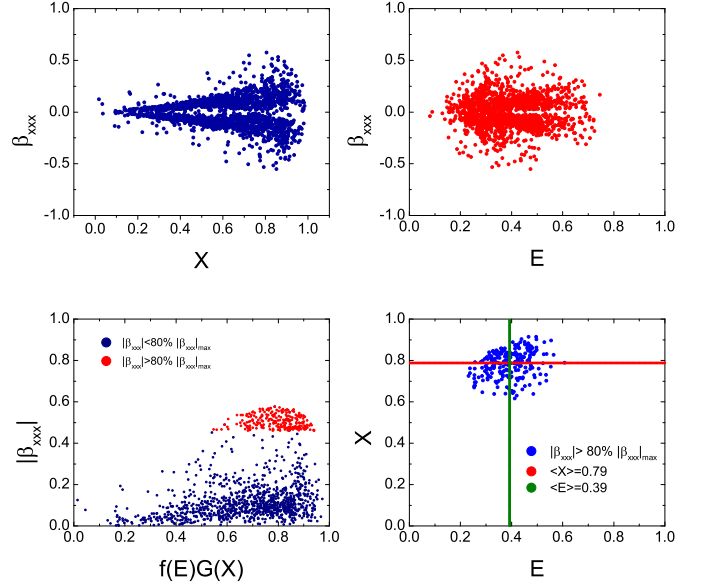}
\caption{(Color online) Dependence of $\beta$ on X and E for a 3-star graph.  The values of X and E when $\beta_{xxx}$ approaches its maximum value are universal for star geometries and for topologies containing star vertices.}
\label{fig:3XEfG}
\end{figure}

Figure \ref{fig:3XEfG} illustrates the range of variation of $\beta$ with E and X (top two panels), and the convergence of $\beta$ to the universal values shown in Fig. \ref{fig:XandE} and expected using the three-state model. This behavior illustrates how the single-electron model achieves universal results in spite of its simplicity, indicating that the dominant state and energy spectrum features of the models are reflective of universal properties of real quantum systems with a Hamiltonian satisfying the sum rules.

While the properties of quantum graphs presented here provide insights into the effects of topology and geometry on the nonlinear optical response, as well as providing a test of the three-level ansatz and universal properties, one might argue that this is just a toy model that does not represent any real system.  The reader must keep in mind that the Monte Carlo approach samples a large configuration space, so the important question is whether this approach catches the important geometrical and topological properties shared by all systems.  While it is difficult to answer this question with rigorous proof, there are many examples where the universal behavior is observed to be independent of the details of the model.  In the case of potential optimization, the universal properties are the same for both one electron, and two interacting electron systems.  As such, our work should be viewed as providing a broad picture of the new possibilities.

\section{Conclusion and outlook}\label{sec:Conclusion}

We have presented for the first time an exact, quantitative dependence of the nonlinear optics of quantum graphs on their geometry and topology.  We have shown that the effects of the topology of geometrically similar graphs dominate those of the geometry of topologically similar graphs.  Topology largely determines the eigenstates and spectra, whereas the geometry mainly affects the projections of the graph edges in the $x-y$ plane.  Closed loop graphs always have non-optimum $\beta_{xxx} $ and negative $\gamma_{xxxx} $, but opening a vertex immediately raises $\beta_{xxx} $ by over a factor of three and enables graphs with positive $\gamma_{xxxx} $. We have also verified that additional degrees of freedom enhance the nonlinearity of the graph, unless a fundamental topological constraint is in place, such as a closed loop boundary condition on the eigenstates.  Finally, we have shown that the star graph has the largest intrinsic hyperpolarizability calculated to date in a simple quantum system.

These results arise from our quasi-one dimensional, one electron free particle model of dynamics on a quantum graph. This useful model enables precision computations of very large numbers of representative graphs of a given topology for many topologies of practical and theoretical interest \cite{shafe12.01}.  It is natural to speculate on the implications of this model that should carry through to more complete models of quantum confined systems that include multiple electrons, band-filling, leaky transverse modes, and cross-coupling between longitudinal and transverse modes at a vertex.  The separation of the nonlinear response into angular and mode-dependent factors is a universal property of the sum-over-states and should remain valid for all graphs. The single-electron model also captures the dominant state and energy spectrum features of other model quantum systems, including their universal scaling properties as the nonlinearities approach their maximum theoretical limits \cite{shafe11.01}.

In our view, the major (and perhaps most interesting) physics will be in systems with very large numbers of quantum states and several length (energy) scales, as we may anticipate the appearance of interesting new behavior for the nonlinearities. But we are likely to retain the same dependence on geometry across topologies, and on topologies for a given geometry.

\begin{acknowledgments}
S. Shafei, J. Smith and M. G. Kuzyk thank the National Science Foundation (ECCS-1128076) for generously supporting this work.
\end{acknowledgments}


%

\end{document}